\documentclass[prb,twocolumn,showpacs,preprintnumbers,superscriptaddress]{revtex4}

\usepackage{amsmath}
\usepackage{amssymb}
\usepackage{graphicx}

\newcommand{\sgn}{\mathop{\rm sgn}\nolimits}
\newcommand{\diag}{\mathop{\rm diag}\nolimits}
\newcommand{\cotanh}{\mathop{\rm cotanh}\nolimits}
\renewcommand{\Re}{\mathop{\rm Re}\nolimits}

\newcommand{\erf}{\mathop{\rm erf}\nolimits}
\newcommand{\Res}{\mathop{\rm Res}}
\newcommand{\Tr}{\mathop{\rm Tr}\nolimits}
\newcommand{\aCom}[2]{{\{ #1, #2 \}}}
\newcommand{\Com}[2]{{[ #1, #2 ]}}
\newcommand{\dg}{\dagger}

\newcommand{\hti}{{\tilde h}}
\newcommand{\fti}{a}

\newcommand{\ga}{{\alpha}}
\newcommand{\go}{{\omega}}
\newcommand{\gx}{{\chi}}

\newcommand{\gd}{{\delta}}

\newcommand{\gL}{{\Lambda}}
\newcommand{\gD}{{\Delta}}
\newcommand{\gS}{{\Sigma}}

\newcommand{\gk}{{\kappa}}
\newcommand{\gt}{{\tau}}

\newcommand{\vG}{{\check G}}

\newcommand{\vgt}{{\check \tau}}
\newcommand{\vOne}{{\check 1}}

\newcommand{\cC}{{\cal  C}}
\newcommand{\cS}{{\cal  S}}
\newcommand{\cE}{{\cal E}}
\newcommand{\cP}{{\cal  P}}
\newcommand{\Te}{{T_e}}

\newcommand{\Vb}{{\bar V}}

\newcommand{\BF}{\mathbf}

\begin{document}


\title{Elementary Charge Transfer Processes in Mesoscopic Conductors}

\author{Mihajlo Vanevi\' c}%
\affiliation{Departement Physik, Universit{\" a}t Basel,
Klingelbergstrasse 82, CH-4056 Basel, Switzerland}

\author{Yuli V. Nazarov}
\affiliation{Kavli Institute of Nanoscience, Delft University of
Technology, 2628 CJ Delft, The Netherlands}

\author{Wolfgang Belzig}
\affiliation{Fachbereich Physik, Universit\" at Konstanz, D-78457
Konstanz, Germany}

\date{\today}

\begin{abstract}
  We determine charge transfer statistics in a quantum conductor
  driven by a time-dependent voltage and identify the elementary
  transport processes.
  At zero temperature unidirectional and bidirectional
  single charge transfers occur. The unidirectional processes
  involve electrons injected from the source terminal due to
  excess dc bias voltage. The bidirectional processes involve
  electron-hole pairs created by time-dependent voltage bias. This
  interpretation is further supported by the charge transfer
  statistics in a multiterminal beam splitter geometry in which
  injected electrons and holes can be partitioned into different
  outgoing terminals. The probabilities of elementary processes
  can be probed by noise measurements: the unidirectional
  processes set the dc noise level while bidirectional ones give
  rise to the excess noise. For ac voltage drive, the noise
  oscillates with increasing the driving amplitude.
  The decomposition of the noise into the contributions of
  elementary processes identifies the origin of these oscillations:
  the number of electron-hole pairs generated per cycle increases
  with increasing the amplitude.
  The decomposition of the noise into elementary processes is
  studied for different time-dependent voltages.
  The method we use is also suitable for systematic calculation of
  higher-order current correlators at finite temperature.
  We obtain current noise power and the third cumulant in the
  presence of time-dependent voltage drive.
  The charge transfer statistics at finite temperature can be
  interpreted in terms of multiple charge transfers with
  probabilities which depend on energy and temperature.
\end{abstract}

\pacs{72.70.+m, 72.10.Bg, 73.23.-b, 05.40.-a}



\maketitle

\section{Introduction}

The charge transmitted through a mesoscopic conductor during
a fixed time interval fluctuates because of charge discreteness
and stochastic nature of transport.
The objective of the statistical theory of quantum transport, full
counting statistics, is to completely characterize the probability
distribution of transferred charge.
The field of full counting statistics has attracted significant
attention 
because it provides the most detailed
information on charge transfer accessible in the measurements of
the average current, current noise
power,\cite{art:BlanterButtikerPHYSREP336-00} 
and higher-order current correlations.%
\cite{art:ReuletSenzierProberPRL91-03,art:ReuletCONDMAT05,%
art:BomzeGershonShovkunLevitovReznikovPRL95-05}

The problem of full counting statistics has been
addressed first by Levitov and Lesovik%
\cite{art:LevitovLesovikJETPLett58-93,art:LevitovLesovikCONDMAT94}
for the case of dc biased multiterminal junctions and subsequently
generalized to a time-dependent voltage bias.%
\cite{art:IvanovLevitovJETPLett58-93,%
art:LevitovLeeLesovikJMathPhys37-96}
In addition to the scattering theory of Levitov and Lesovik
(see also
Ref.~\onlinecite{art:HasslerSuslovGrafLebedevLesovikBlatterCONDMAT08}),
the theoretical approaches to full counting statistics include
the so-called stochastic path-integral approach%
\cite{art:PilgramJordanSukhorukovBuettikerPRL90-03,%
art:JordanSukhorukovPilgramJMathPhys45-04}
and the quantum-mechanical theory
based on an extension of the Keldysh-Green's function technique.%
\cite{art:NazarovAnnPhys8SI193-99,art:BelzigNazarovPRL87-01,%
art:BelzigNazarovPRL87a-01,art:SamuelssonBelzigNazarovPRL92-04,%
art:SnymanNazarovCONDMAT08}
Although equivalent, these theories provide different
methods to access the charge transfer statistics.
Using the extended Keldysh-Green's functions technique, the theory
of full counting statistics of a general quantum-mechanical
observable has been formulated,\cite{art:NazarovKindermannEPJB35-03,%
art:KindermannNazarovCONDMAT03}
including the effects of a detector back
action.\cite{art:KindermannNazarovBeenakkerPRB69-04}
The problem of quantum measurement in the context of full counting
statistics of non-commuting spin observables has also been
analyzed.\cite{art:diLorenzoNazarovPRL93-04,%
art:diLorenzoCampagnanoNazarovPRB73-06}
The discretization\cite{art:NazarovSUPERLATTMICROSTRUCT25-99,%
art:NazarovGenOhmLawCONDMAT94} in space
of the Keldysh-Green's functions results in a quantum circuit
theory\cite{art:BelzigCONDMAT02,art:BelzigCONDMAT03,%
art:NazarovHTCN01-05}
which greatly simplifies the calculations.
The circuit theory can describe junctions with different connectors
and leads, as well as multiterminal
circuits.\cite{art:NazarovBagretsPRL88-02}


The goal of evaluation of full counting statistics in mesoscopic
conductors has essentially been accomplished: the aforementioned
theories provide methods and techniques to calculate the cumulant
generating function of transferred charge from which the
probability distribution can be obtained. However, the {\it
interpretation} of the resulting statistics is often not
straightforward. This is because the total statistics are composed
of many electrons injected
towards conductor, each exhibiting different chaotic scattering
and multiple reflections before entering an outgoing terminal. At
finite temperatures, the charge flow is bidirectional with both
injection and absorption of electrons from terminals. The particle
correlations due to Pauli principle, interactions, and those
induced by the superconducting and/or ferromagnetic proximity
effect also affect the charge transfer statistics.

To {\it understand} the statistical properties of collective
charge transfer, one has to find the independent elementary
processes constituting the statistics. The independent processes
can be revealed by decomposition of the total cumulant generating
function into a sum of simpler ones.
The physical interpretation obtained this way goes beyond
information contained in the average current, noise,
and finite-order cumulants, and pertains to {\it all} transport
measurements. For example, the elementary processes in a dc biased
normal junction
are single-electron transfers, which are independent at different
energies. At finite temperatures, the electrons are transferred in
both directions with correlated left- and right-transfers. At low
temperatures only unidirectional electron transfers remain, with
direction set by polarity of applied voltage. The elementary
processes in a superconductor/normal contact
are single- and double-charge transfers.%
\cite{art:MuzykantskiiKhmelnitskiiPRB50-94,art:BelzigCONDMAT02}
At low temperatures and voltages below the superconducting gap only
double-charge Andreev processes remain,
while above the gap normal single-charge transfers dominate.
The elementary charge transfer processes between
superconductors with dc bias applied consist of multiple
charge transfers due to multiple Andreev reflections.
\cite{art:CuevasBelzigPRL91-03,art:CuevasBelzigPRB70-04,%
art:JohanssonSamuelssonIngermanPRL91-03,%
art:PilgramSamuelssonPRL94-05}
The interpretation of elementary charge transfer processes between
superconductors with constant phase difference is more subtle.
In this case charge transfers acquire formally negative
probabilities and the proper physical interpretation has to
include dynamics of a detector.
\cite{art:BelzigNazarovPRL87-01,art:NazarovKindermannEPJB35-03,%
art:KindermannNazarovCONDMAT03}



In contrast to dc voltage bias, time-dependent voltage drive mixes
electron states of different energies which in combination with
the Pauli principle leads to a nontrivial charge transfer
statistics.
The signatures of this statistics have been studied first
through the noise in an ac-driven junction (photon-assisted noise).%
\cite{art:LesovikLevitovPRL72-94,art:PedersenButtikerPRB58-98}
The photon-assisted noise at low temperatures is a piecewise linear
function of a dc voltage offset with kinks corresponding to
integer multiples of the driving frequency and slopes
which depend on the amplitude and shape of the ac component.
This dependence has
been observed experimentally in normal coherent conductors%
\cite{art:SchoelkopfKozhevnikovProberRooksPRL80-98,%
art:ReydelletRocheGlattliEtienneJinPRL90-03}
and diffusive normal metal/superconductor junctions.%
\cite{art:KozhevnikovProberPRL84-99}
The photon-assisted noise at small driving amplitudes is due to an
{\it electron-hole pair}
which is created by ac drive (with a low probability per voltage
cycle) and injected towards the scatterer.%
\cite{art:RychkovPolianskiButtikerPRB72-05,%
art:PolianskiSamuelssonBuettikerPRB72-05,art:e-hPairPumpung}
However, to obtain the elementary processes, the knowledge of full
charge transfer statistics is needed.

The elementary processes in the presence of time-dependent drive
have been identified in Refs.~%
\onlinecite{art:LeeLevitovCONDMAT95,art:IvanovLeeLevitovPRB56-97,%
art:KeelingKlichLevitovPRL97-06} for the special choice of a
driving voltage. The authors have studied Lorentzian voltage
pulses of the same sign carrying integer numbers of charge quanta.
In this case the elementary processes are electrons injected
towards the scatterer without creation of the electron-hole pairs.
The charge transfer is unidirectional with binomial statistics set
by the effective dc voltage. This result is independent of the
relative position of the pulses, their duration, and overlap. The
many-body quantum state generated by these pulses has been
obtained recently.\cite{art:KeelingKlichLevitovPRL97-06}
An alternative way to inject single electrons free of
electron-hole pairs is by using time-dependent shifts of a
resonant level in a quantum
dot.\cite{art:KeelingShytovLevitovCONDMAT08,%
art:FeveMaheBerroirKontosGlattliEtienneSCIENCE316-07,%
art:MaheFeveKontosGlattliPlacaisCONDMAT08,%
art:MoskaletsSamuelssonButtikerPRL100-08,%
art:OlkhovskayaSplettstoesserMoskaletsButtikerCONDMAT08}

The elementary charge transfer processes for arbitrary
time-dependent voltage drive have been identified in
Ref.~\onlinecite{art:VanevicNazarovBelzigPRL99-07}.
At low temperatures these processes are single-charge transfers
which originate from electrons and electron-hole pairs injected
towards the scatterer. The electrons are injected due to excess
dc voltage applied and give a binomial contribution to
the total cumulant generating function. The electron-hole pairs
are created by the ac component of the voltage drive. The
probabilities of pair creations per voltage cycle depend on the
details of the ac drive. For the special choice of optimal
Lorentzian voltage pulses no electron-hole pairs are created.
In general, however, an ac drive does create electron-hole pairs,
with more and more pairs created per voltage cycle as the
amplitude of the drive increases. A geometric interpretation of
elementary processes is studied in
Ref.~\onlinecite{art:SherkunovPratapMuzykantskiiAmbrumenilCONDMAT08}.
For {\it time-dependent scatterer}, the constraints imposed on
charge transfer statistics restrict the allowed charge
transfers.\cite{art:AbanovIvanovPRL100-08}

\begin{figure}[t]
\includegraphics[scale=1.2]{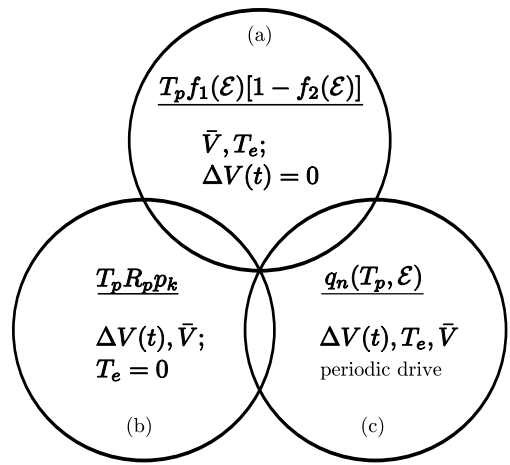}
\caption{\label{fig:ElemProcDiagram}
    The elementary processes shown in different
    regimes. For a dc bias voltage $\Vb$ and at arbitrary
    temperature $\Te$, the electrons are transferred independently
    at different energies. The probability of transfer is given by
    occupation numbers $f_{1,2}(\cE)$ of the leads and
    transmissions $\{T_p\}$ of the junction.
    The transport occurs when the incoming electronic state
    is occupied, the outgoing state is empty, and the electron
    is transferred across the junction (a).
    On the other hand, an ac voltage drive $\gD V(t)$ mixes the
    electron states of different energies. At zero temperature it
    creates the electron-hole pairs with probabilities
    $p_k$ (b), in addition to electrons injected due to excess dc
    voltage. A created electron-hole pair contributes to transport
    only if one particle from a pair is transmitted and the other is
    reflected, which occurs with probability $T_pR_p$ ($R_p$
    being reflection coefficient). Finally, at finite temperature
    and in the presence of a time-dependent voltage drive, the
    statistics can be interpreted in terms of multiple $n$-charge
    transfers with probabilities $q_n$ (c).
}
\end{figure}

In this article we present a comprehensive study of elementary
processes in a generic mesoscopic conductor in different physical
regimes, as depicted in Fig.~\ref{fig:ElemProcDiagram}.
We first obtain the cumulant generating function for a coherent
voltage-driven quantum contact at finite temperatures, using both
the formalism of Levitov and Lesovik and the circuit theory of
quantum transport (Sec. \ref{sec:CGF}). We then identify the
elementary charge transfer processes {\it at zero temperature}
(Sec.~\ref{sec:ElemProcTeq0-decomp}). Decomposition of the
noise into contributions of elementary processes for different
voltage signals is studied in Sec.~\ref{sec:CompareVt}.
For ac voltage drive, the differential noise oscillates as the
amplitude of the drive increases due to new electron-hole pairs
being created per voltage cycle.
The number of electron-hole pairs per cycle becomes large
for driving amplitudes much larger than frequency.
In this case the statistics in the leading order reduces to
uncorrelated electron and hole transfers and depends on an
effective voltage only, independent of the details of the
ac drive. The elementary processes can be probed also
for both ac and dc voltages present, e.g., in the
regime in which the ac component of the drive is kept fixed
and the dc offset changes. In this case both the electrons
and created electron-hole pairs are injected towards the
junction. The electron-hole pairs give rise to the excess noise
with respect to the noise level set by dc voltage offset.

The notion of elementary processes, being a matter of interpretation
of the full charge transfer statistics, provides physically
plausible and intuitive description of transport in terms of
statistically independent events. Such a description persists in
all transport measurements. For example, the picture of electrons
and electron-hole pairs injected towards the scatterer, which we
have inferred from a study of a 2-terminal junction, is further
supported by the evaluation of the full counting statistics in
a beam-splitter geometry in which case the injected particles can
be separated into different outgoing terminals.
These processes can be probed by current cross correlations
(Sec.~\ref{sec:BeamSplitter}).

The method we use is also suitable for systematic calculation of the
higher-order current correlators {\it at finite temperature}. The
current noise power and the third cumulant in the presence of
time-dependent voltage drive are obtained in
Sec.~\ref{sec:CumulantsTneq0}.
The interpretation of the full counting statistics and the
elementary processes are fundamentally different at finite
temperature. For a periodic drive, the elementary processes are
transfers of multiple integer charge quanta with probabilities which
depend on energy and temperature (Sec.~\ref{sec:ElemProcTne0}).



\section{Cumulant generating function}
\label{sec:CGF}

The system we consider is a generic 2-terminal mesoscopic conductor
characterized by a set of transmission eigenvalues $\{T_p\}$.
For a large number of channels, the transport properties are
universal and independent of microscopic details of geometry
of the junction and positions of impurities.
We can neglect the energy dependence of transmission eigenvalues
if the the electron dwell time is small with respect to time
scales set by the inverse temperature and applied voltage.
%
The cumulant generating function of the transferred charge
is given by determinant formula%
\cite{art:LevitovLesovikJETPLett58-93,art:IvanovLeeLevitovPRB56-97,%
art:CommentDetFormula}%
\begin{equation}\label{eq:LevitovFla}
\cS(\gx) = \ln \det
\left[
\BF 1 + \BF f
    \left(
    \BF S^\dg \BF \gL^\dg_\gx \BF S \BF \gL_\gx - \BF 1
    \right)
\right],
\end{equation}
where
\begin{equation}
\BF f = \begin{pmatrix}
        f_1 & 0 \\
        0 & f_2
        \end{pmatrix},
\quad
\BF S = \begin{pmatrix}
        \BF r & \BF t' \\
        \BF t & \BF r'
        \end{pmatrix},
\quad
\BF \gL_{\gx} =
        \begin{pmatrix}
        e^{-i\gx} & 0 \\
        0 &  1
        \end{pmatrix}.
\end{equation}
Here $\BF f$ is the matrix of occupation numbers of the terminals
which is diagonal in the terminal indices and scalar in transport
channels, $\BF S$ is the scattering matrix of the junction,
and $\BF{\gL}^\dg_\gx \ldots \BF{\gL}_\gx$ is the
transformation which incorporates the counting fields.
Because the current is conserved, it is sufficient to
assign only one counting field $\gx$ to the left terminal.
We count charges which enter the left terminal irrespective the
channel, the energy, or the spin, 
with $\gx$ being scalar in the corresponding indices.
The determinant in Eq.~\eqref{eq:LevitovFla} is taken with respect
to the terminal, channel, energy, and spin indices.

The occupation numbers $f_1$ and $f_2$ of the left and right
terminals are matrices in energy indices and scalars in channel
and spin indices.
In the case of dc voltage bias $f_{1,2}$ are diagonal in energy with
$f_i(\cE',\cE'') = f_i(\cE') \; 2\pi\gd(\cE'-\cE'')$,
where $f_i(\cE)=(e^{(\cE-eV_i)/T_e}+1)^{-1}$,
$V_i$ is the voltage applied, and $\Te$ is the electronic
temperature. In the presence of time-dependent drive the
occupation numbers $f_i(\cE',\cE'')$ are not diagonal in energy
indices and do not commute. The voltage drive $V(t)$ can be
incorporated via the gauge transformation in time representation
$f_1 \to U f_{V=0} U^\dg$ with $U(t',t'')=
e^{-i \int_0^{t'}eV(t) dt} \gd(t'-t'')$, where we assume the
convolution over internal time indices.

Equation~\eqref{eq:LevitovFla} can be simplified using
polar decomposition of the scattering
matrix~\cite{art:BeenakkerRMP69-97}
\begin{equation}
\BF S = \begin{pmatrix}
        \BF U & \BF 0 \\
        \BF 0 & \BF V
        \end{pmatrix}
        \; \BF S' \;
        \begin{pmatrix}
        \BF U' & \BF 0 \\
        \BF 0 & \BF V'
        \end{pmatrix},
\quad
\BF S'= \begin{pmatrix}
        -\sqrt{\BF R} & \sqrt{\BF T} \\
        \sqrt{\BF T} & \sqrt{\BF R} \\
        \end{pmatrix}.
\end{equation}
Here $\BF U$, $\BF V$, $\BF U'$, and $\BF V'$ are unitary
matrices in transport channels and $\BF T = \diag(T_1,T_2,\ldots)$
and $\BF R = \BF 1 - \BF T$ are diagonal matrices of transmission
and reflection eigenvalues. Equation~\eqref{eq:LevitovFla}
for the 2-terminal junction reduces to
\begin{multline}
\cS(\gx)
=
2_s \ln \det
\left[ \BF 1 + \BF f
    \left(
    \BF S' \BF{\gL}^\dg \BF S' \BF{\gL} - \BF 1
    \right)
\right]
\\
=
2_s \ln\det
    \begin{pmatrix}
    1 + f_1\BF T (e^{-i\gx}-1) & -f_1 \sqrt{\BF{TR}} (e^{i\gx}-1) \\
    f_2 \sqrt{\BF{TR}}(e^{-i\gx}-1) & 1 + f_2 \BF T (e^{i\gx}-1)
    \end{pmatrix}
,
\end{multline}
where $2_s$ takes into account the spin degree of freedom.
Since the operators in the first (the second) row commute, the
determinant can be calculated blockwise. Using
Eq.~\eqref{eq:DetBlockMatr} given in Appendix we obtain
\begin{widetext}
\begin{subequations}\label{eq:LevitovCGF2terminal}
\begin{align}\label{eq:LevitovCGF2terminal1}
\cS(\gx)
=&
2_s \sum_p \Tr_\cE \ln
[ 1 + (1-f_1)f_2 T_p (e^{i\gx}-1) + f_1(1-f_2) T_p (e^{-i\gx}-1)]
\\
\label{eq:LevitovCGF2terminal2}
=&
2_s \sum_p \Tr_\cE \ln
[1 + f_2(1-f_1) T_p (e^{i\gx}-1) + (1-f_2)f_1 T_p (e^{-i\gx}-1)].
\end{align}
\end{subequations}
\end{widetext}
Here we used the matrix identity
$\ln \det ({\BF M}) = \Tr\ln ({\BF M})$ and the trace $\Tr_\cE$
is taken in energy indices. The logarithm is taken assuming
convolution over internal energy indices, e.g.,
$(f_1 f_2)_{\cE'\cE''} = (2\pi)^{-1}
\int d\cE_1\; f_1(\cE',\cE_1)f_2(\cE_1,\cE'')$.

\begin{figure}[b]
\includegraphics[scale=0.95]{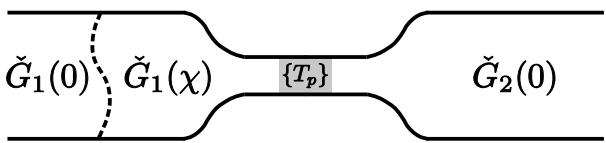}
\caption{\label{fig:TwoTerminalJunction}
  The schematic circuit-theory representation of
  a 2-terminal junction. The left and right terminals are
  characterized by the quasiclassical Keldysh-Green's functions
  $\vG_{1,2}(0)$. The counting field $\gx$ is related to the
  charges which enter the left terminal through the cross-section
  indicated by dashed line. It can be incorporated into the
  corresponding Keldysh-Green's function via gauge transformation
  given by Eq.~\eqref{eq:vG1chi}.}
\end{figure}

In the following we show that the same result is obtained
within the circuit theory of mesoscopic transport.%
\cite{art:BelzigCONDMAT02,art:BelzigCONDMAT03,art:NazarovHTCN01-05}
In this case the system is represented by discrete circuit elements
as depicted in Fig.~\ref{fig:TwoTerminalJunction}.
The cumulant generating function of the
transferred charge is given by%
\cite{art:NazarovSUPERLATTMICROSTRUCT25-99,%
art:SnymanNazarovCONDMAT08}
\begin{equation}\label{eq:Schi-CircuitTh}
\cS(\gx) = \sum_p \Tr \ln
\left[
    \vOne + \frac{T_p}{2}
    \left(
    \frac{\aCom{\vG_1(\gx)}{\vG_2(0)}}{2} - \vOne
    \right)
\right],
\end{equation}
where
\begin{equation}
\vG_i(0) =
    \begin{pmatrix}
    1 & 2h_i \\
    0 & -1
    \end{pmatrix}
\end{equation}
are the quasiclassical Keldysh-Green's functions of the terminals
with $h_i = 1-2f_i$. The counting field is incorporated through the
gauge transformation of the Keldysh-Green's function of the
left terminal,
\begin{equation}\label{eq:vG1chi}
\vG_1(\gx) = e^{-i\gx \vgt_1/2}\;\vG_1(0)\;e^{i\gx \vgt_1/2},
\end{equation}
where
\begin{equation}
\vgt_1 =\begin{pmatrix}
        0 & 1 \\
        1 & 0
        \end{pmatrix}
\end{equation}
in Keldysh space.
The logarithm and the trace in Eq.~\eqref{eq:Schi-CircuitTh} are
taken assuming the convolution both in Keldysh and in energy
indices.

Equation~\eqref{eq:Schi-CircuitTh} can be simplified by
using matrix representation in Keldysh space. After rewriting
$\Tr\ln(\cdots)=\ln \det(\cdots)$ and taking  the determinant
by blocks using Eq.~\eqref{eq:DetBlockMatr}, the result
coincides with Eq.~\eqref{eq:LevitovCGF2terminal}.
This proves the equivalence of the circuit-theory expression
for $\cS(\gx)$ [Eq.~\eqref{eq:Schi-CircuitTh}]
and the Levitov determinant formula
[Eq.~\eqref{eq:LevitovFla}] for
a coherent 2-terminal scatterer. The circuit-theory approach
becomes advantageous in the case of several scatterers in series%
\cite{art:NazarovGenOhmLawCONDMAT94,art:VanevicBelzigEPL75-06} or
for a multiterminal mesoscopic conductor with large number of
conduction channels.\cite{art:NazarovBagretsPRL88-02}
It gives the charge transfer statistics in
terms of the scattering properties of individual elements, thus
circumventing a non-trivial task of obtaining the scattering
matrix of the composite system and averaging over the phase
shifts.

Equation~\eqref{eq:LevitovCGF2terminal} is valid for a dc bias
applied and energy-dependent transmission probabilities. In this
case the occupation numbers $f_{1,2}$ are diagonal in energy and
commute with each other and with $T_p(\cE)$. The trace over energy
reduces simply to the integration and we obtain
\begin{multline}\label{eq:CGFdc}
\cS(\gx) =
\frac{t_0}{\pi} \sum_p\int d\cE\; \ln
\big(
1 + [1-f_1(\cE)]f_2(\cE) T_p (e^{i\gx}-1)
\\
+ f_1(\cE)[1-f_2(\cE)]
T_p (e^{-i\gx}-1)
\big).
\end{multline}
Here $t_0$ is the measurement time which is the largest time
scale in the problem, much larger than the characteristic time
scale on which the current fluctuations are correlated.%
\cite{art:CommentFiniteTime}
The form of $\cS(\gx)$ reveals that the
elementary charge transfer processes are single-electron transfers
to the left and to the right. The electron
transfers at different energies and in different channels are
independent. The term $[1-f_1(\cE)]f_2(\cE)T_p e^{i\gx}$ in
Eq.~\eqref{eq:CGFdc} describes the electron transfer from the
right to the left lead at energy $\cE$ in the channel $p$.
The probability of this process is proportional to the
probability $f_2(\cE)$ that the state in the right lead is
occupied, the probability $1-f_1(\cE)$ that the state in the left
lead is empty, and the probability $T_p(\cE)$ of transfer across
the scatterer. A similar analysis holds for the electron transfer
from left to right. The left and right transfers are correlated
because $\cS(\gx)$ is not a sum of the left- and right-transfer
generating functions.

Equation~\eqref{eq:LevitovCGF2terminal} is also valid for the
time-dependent voltage applied and {\it energy-independent}
transmission probabilities.
In this case $f_{1,2}(\cE',\cE'')$ do not commute
with each other.
[However, the order of $f_1$ and $f_2$
can be exchanged as shown by Eqs.~\eqref{eq:LevitovCGF2terminal1}
and~\eqref{eq:LevitovCGF2terminal2}.] %
The logarithm has to be calculated with the matrix structure of
$f_{1,2}$ in energy indices taken into account because
time-dependent drive mixes the electron states with different
energies.

A few remarks on applicability of our approach are in place here.
The circuit theory we have used applies to instantaneous scattering
at the junction, with the frequency $\go$ of the bias
voltage much smaller than the inverse dwell time $\gt_{\rm d}^{-1}$.
It also applies for a dot sandwiched between 2 terminals and
$\go$ much smaller than the inverse $RC$ time
$\gt_{RC}^{-1}$ of the charge relaxation on the dot.
In a typical experimental situation
$\gt_{RC}\ll \gt_{\rm d}$,%
\cite{art:BagretsPistolesiPRB75-07,art:BagretsPistolesiPhysicaE40-07}
with $\go$ limited by the dwell time $\go\ll \gt_{\rm d}^{-1}$.
However, as shown in Ref.
\onlinecite{art:PolianskiSamuelssonBuettikerPRB72-05},
the photon-assisted
noise $S_I(\go)$ in the leading order $(eV_0/\go)^2\ll 1$
in driving amplitude $V_0$ does not
depend on $\gt_{\rm d}$:
in a {\it weakly-driven}
junction the frequency range is set by the $RC$ time,
$\go \ll \gt_{RC}^{-1}$.
The dwell time becomes important for larger voltage amplitudes.
For $eV_0/\go \gtrsim 1$ a maximum in $\partial S_I/\partial \go$
develops at the frequency $\go\sim \gt_{\rm d}^{-1}$
because the photon-assisted noise $S_I(\go)$ probes the
{\it electronic distribution function} $f(\cE; t) =
\int d\gt\; f(t+\gt/2,t-\gt/2)\; e^{i\cE\gt}$ which relaxes on a
timescale given by $\gt_{\rm d}$.%
\cite{art:BagretsPistolesiPRB75-07,art:BagretsPistolesiPhysicaE40-07}
At frequencies $\go\sim\gt_{\rm d}^{-1}$, the differential noise
$\partial S_I/\partial V_0$ as a function of $V_0$
is increased with respect to the case
$\go\ll \gt_{\rm d}^{-1}$.
The increase is of the order of $20\%$ with the distances between
maxima and minima only weakly affected. This is consistent
with the experiments of
Schoelkopf {\it et al.}
\cite{art:SchoelkopfKozhevnikovProberRooksPRL80-98}
and
Kozhevnikov {\it et al.}
\cite{art:KozhevnikovProberPRL84-99} which can be described by
assuming the instantaneous scattering even though
the frequency of the ac signal applied is comparable or larger
than the inverse dwell time.

\section{Elementary charge transfer processes at $\Te=0$}
\label{sec:ElemProcTeq0}

\subsection{Decomposition of the cumulant generating function into
elementary processes}
\label{sec:ElemProcTeq0-decomp}

In this section we study elementary charge transfer processes in
a 2-terminal junction with constant transmission eigenvalues
$\{T_p\}$ and time-dependent voltage $V(t)$ applied. The cumulant
generating function $\cS(\gx)$ is given by
Eq.~\eqref{eq:LevitovCGF2terminal}. To identify the elementary
processes we diagonalize the operator under the
logarithm in energy indices. As a first step, we rewrite
Eq.~\eqref{eq:LevitovCGF2terminal2} in terms
of $\hti \equiv h_1$ and $h \equiv h_2$:
\begin{multline}\label{eq:Schi-h}
\cS(\gx) = 2 \sum_p \Tr_\cE \ln
[1-T_p(1-h\hti) \sin^2(\gx/2)
\\
-i T_p(h-\hti)\sin(\gx/2)\cos(\gx/2)],
\end{multline}
where $h(\cE',\cE'')=\tanh(\cE'/2\Te) 2\pi\gd(\cE'-\cE'')$ and
$\hti=UhU^\dg$.\cite{art:CommentCGFeq}
%
The $\cS(\gx)$ further simplifies in the zero
temperature limit, in which the hermitian $h$-operators are
involutive $h^2=\hti^2=1$, and satisfy
$\Com{h}{\aCom{h}{\hti}}=\Com{\hti}{\aCom{h}{\hti}}=0$. Thus
the eigensubspaces of $\aCom{h}{\hti}$ are invariant with respect
to $h$, $\hti$, and $h\hti$, and the diagonalization problem
reduces to the subspaces of lower dimension.
The {\it typical} eigensubspaces of
$\aCom{h}{\hti}$ are 2-dimensional and spanned by the eigenvectors
$\BF v_\ga$ and $\BF v_{-\ga}=h\BF v_\ga$ of $h\hti$ which
correspond to the eigenvalues $e^{\pm i\ga}$ ($\ga$ is real).
The diagonalization procedure in invariant subspaces is described
in detail in Ref.~\onlinecite{art:VanevicNazarovBelzigPRL99-07}.

For computational reasons it is convenient to impose periodic
boundary conditions on the voltage drive $V(t+\gt)=V(t)$
with the period $\gt=2\pi/\go$. In this case the operator
$\hti=UhU^\dg$ couples
only energies which differ by an integer multiple of $\go$. This
allows to map the energy indices into the interval
$0<\cE<\go$ while retaining the discrete matrix structure in steps
of $\go$. The operator $h\hti$ in energy representation is given
by
\begin{multline}
(h\hti)_{nm}(\cE) \equiv (h\hti)(\cE+n\go,\cE+m\go)
\\
=
\sgn(\cE+n\go) \sum_{k=-\infty}^\infty \fti_{n+k} \fti^*_{m+k}
\sgn(\cE-k\go-e\Vb),
\end{multline}
with
\begin{equation}\label{eq:fti-def}
\fti_n = \frac{1}{\gt}\int_0^\gt dt\;
e^{-i\int_0^t dt' e\gD V(t')}e^{in\go t}.
\end{equation}
Here $\Vb=(1/\gt)\int V(t)dt$ is the dc voltage offset and
$\gD V(t)=V(t)-\Vb$ is the ac voltage component. The coefficients
$\fti_n$ satisfy
\begin{equation}\label{eq:fti-prop}
\sum_{k=-\infty}^\infty \fti_{n+k}\fti^*_{m+k} = \gd_{nm}
\quad
\text{and}
\quad
\sum_{n=-\infty}^\infty n|\fti_n|^2 = 0.
\end{equation}

The cumulant generating function $\cS(\gx)$ at zero temperature
is given by $\cS = \cS_{1L}+ \cS_{1R} + \cS_2$
with\cite{art:VanevicNazarovBelzigPRL99-07}
\begin{multline}\label{eq:S1LR}
\cS_{1L,R}(\gx)
= M_{L,R}
\\
\times
\sum_n \sum_k
\ln \left[ 1 + T_nR_n p_{kL,R}
        \left( e^{i\gx} + e^{-i\gx} -2 \right)
    \right],
\end{multline}
and
\begin{equation}\label{eq:S2}
\cS_2(\gx) = \frac{t_0 |e\bar V|}{\pi} \sum_n
\ln \left[ 1+T_n \left( e^{-i\gk \gx} - 1 \right)\right].
\end{equation}
Here $M_L = t_0\go_1/\pi$, $M_R = t_0(\go-\go_1)/\pi$, and
$\go_1=e\Vb - \lfloor e\Vb/\go \rfloor \go$, where
$\lfloor x \rfloor$ is the largest integer less than
or equal $x$.
The coefficient $\gk=\pm 1$ in Eq.~\eqref{eq:S2} is related to
direction of the charge transfer, with $\gk=1$ ($\gk=-1$) for
$e\Vb>0$ ($e\Vb<0$). The total measurement time $t_0$ is much
larger than the period $\gt$ and the characteristic time scale on
which the current fluctuations are correlated.

The parameters $p_{kL,R}$ in Eq.~\eqref{eq:S1LR} depend on the
details of the time dependent voltage drive. They are given by
$p_{kL,R} = \sin^2(\ga_{kL,R}/2)$, where $e^{\pm i \ga_{kL(R)}}$
are the eigenvalues of $(h\hti)_{nm}$ calculated for
$\cE\in (0,\go_1)$ [$\cE\in(\go_1,\go)$].
The eigenvalues are obtained using a finite-dimensional matrix
$(h\hti)_{nm}$, with the cutoff in indices $n$ and $m$ being much
larger than the characteristic scale on which $|\fti_n|$ vanish.

Equations \eqref{eq:S1LR} and \eqref{eq:S2} give the charge
transfer statistics in a 2-terminal junction driven by a
time-dependent voltage. The result is valid in the zero-temperature
limit, with no thermal excitations present.
The probability distribution of the number of charges
$N$ transferred within measurement time is given
by $\cP(N)= (2\pi)^{-1}\int_{-\pi}^\pi d\gx
\;\exp[\cS(\gx)-iN\gx]$. The cumulants of $N$ are given by $\cC_n
= [\partial_{i\gx}^n \cS(\gx)]_{\gx=0}$ and are related to
the higher-order current correlators at zero frequency.%
\cite{art:NazarovKindermannEPJB35-03,art:KindermannNazarovCONDMAT03}
For example, the average current, the current noise power, and
the third cumulant of current fluctuations are given by
$I = (e/t_0)\partial_{i\gx}\cS|_{\gx=0}$,
$S_I = (e^2/t_0)\partial^2_{i\gx}\cS|_{\gx=0}$,
and $C_I = (e^3/t_0)\partial^3_{i\gx}\cS|_{\gx=0}$,
respectively.\cite{art:SIconvention}

\begin{figure}[t]
\includegraphics[scale=1]{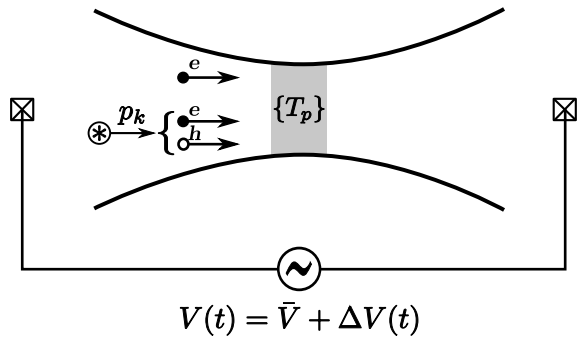}
\caption{\label{fig:ElemProcesses}
    The elementary charge transfer processes in a voltage-driven
    junction. The unidirectional processes represent single electrons
    injected towards the contact due to excess dc offset
    voltage $\Vb$ applied. The bidirectional processes represent
    electron-hole pairs created by time-dependent voltage drive
    and injected towards the contact. The probabilities $p_k$
    of pair creations depend on the ac voltage component $\gD V(t)$.
}
\end{figure}

The elementary charge transfer processes can be identified from the
form of $\cS$, similarly as in
Refs.~\onlinecite{art:TobiskaNazarovPRB72-05} and
\onlinecite{art:diLorenzoNazarovPRL94-05}.
The result is depicted schematically in Fig.~\ref{fig:ElemProcesses}.
The $\cS_2(\gx)$
describes {\it unidirectional} single-electron transfers due to the
excess dc bias voltage $\Vb$ applied. The number of attempts for
an electron to traverse the junction within measurement time
$t_0$ is given by $t_0|e\Vb|/\pi$ (we take into account
both spin orientations). The transfer events in different channels
are independent. The term $T_n e^{-i\gk\gx}$ in Eq.~\eqref{eq:S2}
describes a single-electron transfer with probability $T_n$ in
$n$th channel. The unidirectional charge transfer processes give
contributions to the average current and higher-order cumulants.

The $\cS_{1L}(\gx)$ and $\cS_{1R}(\gx)$ describe {\it bidirectional}
charge transfer processes. Different bidirectional processes are
labelled by $k$ in Eq.~\eqref{eq:S1LR}.
These processes represent electron-hole pairs created
in the source terminal by the time-dependent voltage drive and
injected towards the scatterer. The probability of such an
electron-hole pair creation is given by $p_k$ and depends on the
details of the ac voltage component $\gD V(t)$. The charge
transfer in $n$th channel occurs if one particle (e.g., electron)
is transmitted and the hole is reflected, or vice versa.
The probability for the whole process is given by $T_n R_n p_k$.
The bidirectional processes contribute to the noise and higher-order
even cumulants. However, they give no contributions to the average
current and higher-order odd cumulants because electrons and
holes are transmitted with the same probability.

The two types of bidirectional processes $\cS_{1L}$ and $\cS_{1R}$
differ in the number of attempts $M_{L,R}$
and have different probabilities $p_{kL,R}$.
The $M_{L,R}$ depend on the number $e\Vb/\go$ of
{\it unidirectional} attempts per period per spin. The simplest
statistics is obtained for an integer value of $e\Vb/\go$ for
which $\cS_{1L}$ vanishes, in agreement with
Ref.~\onlinecite{art:IvanovLevitovJETPLett58-93}.

\subsection{Comparison of different time-dependent voltages}
\label{sec:CompareVt}

The elementary processes at zero temperature can be probed by
noise measurements. In what follows we compare the elementary
processes and the noise generated by different time-dependent
voltages. We focus on standard periodic voltage signals such as
cosine, square, triangle, and sawtooth. We also present results
for Lorentzian voltage pulses which provide the simplest charge
transfer statistics for certain amplitudes.
The applied voltage is characterized by coefficients $\fti_n$
given by Eq.~\eqref{eq:fti-def}. These coefficients are calculated
explicitly in Sec.~B 
in Appendix for the driving voltages of interest.

We first consider ac voltage drive with no dc bias applied,
$\Vb=0$. In this case only the bidirectional processes of
R-type remain, $\cS(\gx)=\cS_{1R}(\gx)$.
The number of attempts during the measurement time is given by
$M= 2 t_0/\gt$ which corresponds to a single attempt per
voltage cycle per spin.
The current noise power is given by
\begin{equation}\label{eq:SI-ElemEvents}
S_I = \frac{2e^2\go}{\pi} \left( \sum_p T_pR_p \right)
\left( \sum_k p_k \right),
\end{equation}
with the probabilities $p_k$ of electron-hole pair creations
obtained from the eigenvalues of $(h\hti)_{nm}$, as discussed in
Sec.~\ref{sec:ElemProcTeq0-decomp}.

\begin{figure}[t]
\includegraphics[scale=0.75]{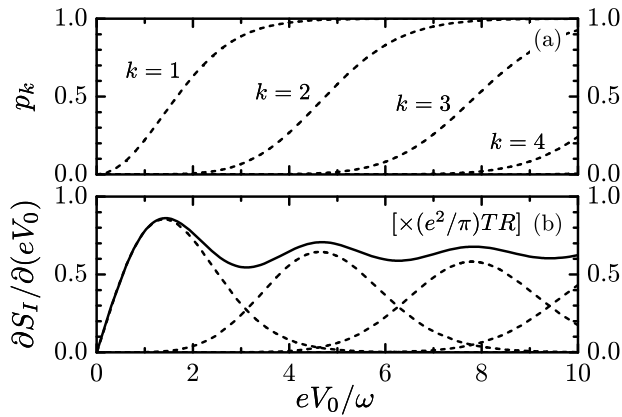}
\caption{\label{fig:ElemProcCos}
    The probabilities $p_k$ of electron-hole pair creations
    for harmonic drive $V(t)=V_0 \cos(\go t)$ as
    a function of the amplitude $V_0$ [panel (a)].
    With increasing $V_0$ more and more
    pairs are created per voltage cycle. The decomposition of
    the differential noise $\partial S_I/\partial (eV_0)$
    (solid line) into contributions of elementary processes
    (dashed lines) is shown in panel (b).
}
\end{figure}

\begin{figure}[t]
\includegraphics[scale=1.6]{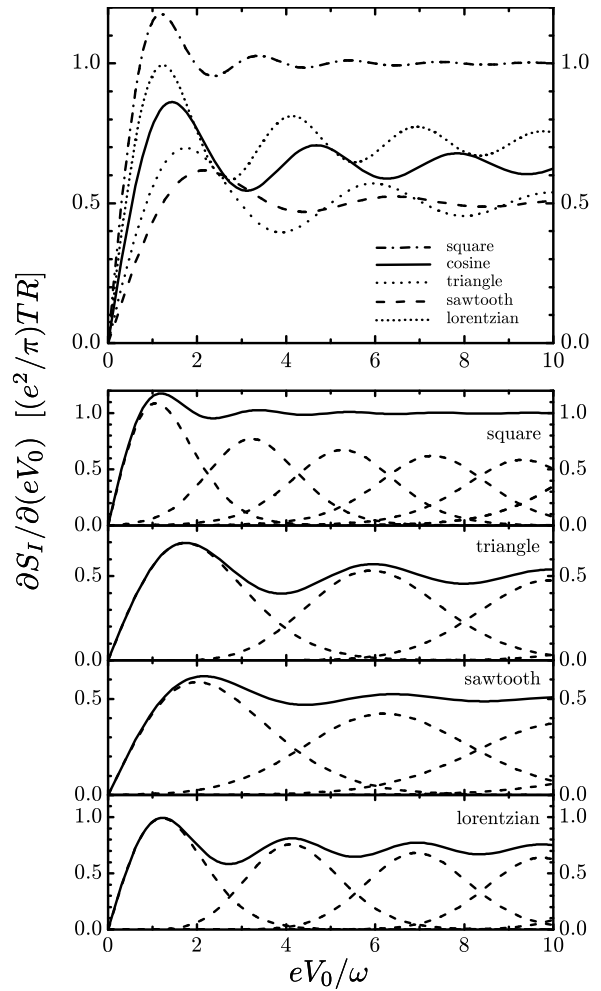}
\caption{\label{fig:NoiseDerivVbEq0All}
    The differential noise $\partial S_I/\partial (eV_0)$ as a
    function of the amplitude $V_0$ for different ac driving
    voltages ($\Vb=0$) is shown in top panel:
    square (dash-dotted), cosine (solid), triangle
    (dotted), sawtooth (dashed), and Lorentzian pulses
    (short-dotted line). The width of Lorentzian pulses is
    $\gt_L=0.1\gt$. The oscillations are due to elementary
    processes which are created as the voltage amplitude increases
    (cf. Fig.~\ref{fig:ElemProcCos}). The decomposition of the
    differential noise into contributions of elementary processes
    is shown in lower panels.
}
\end{figure}

The probabilities $p_k$ for the
harmonic voltage drive $V(t)=V_0 \cos(\go t)$ are shown in
Fig.~\ref{fig:ElemProcCos}(a).
As the amplitude $V_0$ increases the probabilities $p_k$
also increase and new pairs start to enter the transport.
This results in the oscillatory change of the slope of $S_I$
as a function of $V_0$. The decomposition of the differential noise
$\partial S_I/\partial V_0$ into elementary
processes is shown in Fig.~\ref{fig:ElemProcCos}(b).
The differential noise $\partial S_I/\partial V_0$ for different
time-dependent voltages is shown in
Fig.~\ref{fig:NoiseDerivVbEq0All} for comparison.

The bidirectional processes with the unit probability $p_k=1$
represent electron-hole pairs which are {\it created} and injected
towards the scatterer in {\it each} voltage cycle. In this case
the electron and hole transfers are statistically independent.
This can be seen from the corresponding cumulant generating
function which reduces to
$\cS_{nk}= \cS_{nk}^{\rm e} + \cS_{nk}^{\rm h}$, where
$\cS_{nk}^{\rm e,h}(\gx)=(2t_0/\gt) \ln [1+T_n(e^{\mp i\gx}-1)]$
describe electron and hole transport.
The electron-hole pairs which are created with the probability
$0<p_k<1$ result in {\it correlated} electron and hole transfers.
The corresponding cumulant generating function is given by
$\cS_{nk}(\gx)= (2t_0/\gt)\ln [1+T_nR_np_k(e^{i\gx}+e^{-i\gx}-2)]$
and cannot be partitioned into independent electron and
hole contributions.

The interpretation of the shot noise in terms of electron-hole
pair excitations has been studied previously in
Ref.~\onlinecite{art:RychkovPolianskiButtikerPRB72-05} in the
regime of low-amplitude harmonic driving $eV_0/\go \ll 1$.
In this case only one electron-hole pair is
excited per period with probability
$p_1 \approx (e V_0/2\go)^2\ll 1$.
Remarkably, Fig.~\ref{fig:ElemProcCos} shows that the single
electron-hole pair is excited not only for small
amplitudes but also for amplitudes comparable or even larger than
the drive frequency. This extended range of validity can be
covered by taking into account the higher-order terms in the
expression for probability
$p_1\approx \sum_{n=1}^\infty n [J_n(eV_0/\go)]^2$,
where $J_n(x)$ denote Bessel functions
(cf. Sec.~\ref{sec:CumulantsTneq0}).
The first $3$ terms approximate the exact $p_1$
shown in Fig.~\ref{fig:ElemProcCos} to accuracy better than
$0.3\%$ for $eV_0/\go \lesssim 2$.

For large driving amplitudes the charge transfer statistics does
not depend on the details of time-dependent drive%
\cite{art:LevitovLeeLesovikJMathPhys37-96,art:LeeLevitovCONDMAT95}
and can be characterized by an effective voltage
$V_{\rm eff} = \gt^{-1}\int_\gt dt |V(t)|$
(here we assume $\Vb=0$).
In this case there are $N=|eV_{\rm eff}|/2\go \gg 1$ processes
with $p_k=1$. The charge transfer statistics in the leading order
in $N$ consists of $N_{\rm e, \rm h}=2Nt_0/\gt$ uncorrelated
electrons and holes injected towards the contact in each
transport channel during measurement time:
$\cS(\gx) = N_{\rm e} \sum_n \ln[1+T_n(e^{-i\gx}-1)]
+ N_{\rm h} \sum_n \ln[1+T_n(e^{i\gx}-1)] $.
This result can be interpreted by comparison with the
generating function of a dc bias given by Eq.~\eqref{eq:S2}.
The electrons (holes) are injected during time
intervals $\gt_{\rm e}$ ($\gt_{\rm h} = \gt - \gt_{\rm e}$)
per voltage cycle in which $eV(t)>0$ [$eV(t)<0$].
For a large number of injected particles, the time-dependent drive
in these intervals can be replaced by effective dc voltages
$V_{\rm eff}^{({\rm e})}$ and $V_{\rm eff}^{({\rm h})}$.
The number of attempts is given by
$N_i= (t_0/\pi)(\gt_i/\gt) |eV_{\rm eff}^{(i)}|$ ($i=$ e, h),
in agreement with Eq.~\eqref{eq:S2}.
The noise generated is the sum of independent electron and
hole contributions $S_I = (|e^3 V_{\rm eff}|/\pi) \sum_n T_n R_n$.
This explains the asymptotic behavior of differential noise at
large driving amplitudes shown
in Figs.~\ref{fig:ElemProcCos} and \ref{fig:NoiseDerivVbEq0All}.

\begin{figure}[t]
\includegraphics[scale=1.4]{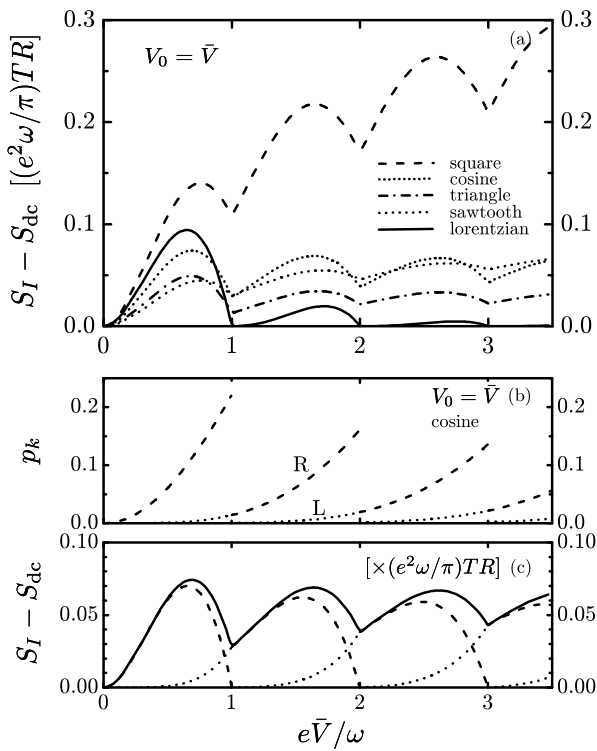}
\caption{\label{fig:ExcessNoiseVbEqV0}
    The excess noise $S_I-S_{\rm dc}$ is shown in panel (a)
    for different time-dependent bias voltages: square (dashed),
    cosine (short-dotted), triangle (dash-dotted), sawtooth
    (dotted), and Lorentzian pulses (solid line).
    The width of Lorentzian pulses is $\gt_L=0.1\gt$.
    The amplitude and the dc offset increase simultaneously,
    $V_0 = \Vb$.
    The probabilities $p_{kL}$ (dotted lines) and $p_{kR}$ (dashed
    lines) of bidirectional elementary processes are shown
    in panel (b) for the cosine voltage drive
    $V(t)=\Vb + \Vb \cos(\go t)$.
    Only one elementary process of the L-type and one of the
    R-type can be created per period ($k=1$). The decomposition of
    the excess noise $S_I-S_{\rm dc}$ into contributions of
    elementary processes is shown in panel (c).
}
\end{figure}
In the following we consider time-dependent voltage drive
with a nonzero dc offset $\Vb\ne 0$ which creates both
unidirectional and bidirectional elementary processes.
The unidirectional processes generate dc noise which is given by
$S_{\rm dc}= (e^2/\pi)(\sum_p T_pR_p)e\Vb$.
The bidirectional processes given by Eq.~\eqref{eq:S1LR} generate
the excess noise $S_I-S_{\rm dc}=(e^2/t_0)
\partial^2_{i\gx}\cS_1|_{\gx=0}$:
\begin{equation}\label{eq:SI-Sdc}
S_I - S_{\rm dc} =
\frac{2e^2\go}{\pi}\left(\sum_p T_pR_p\right)
\sum_k
[ \bar v p_{kL} + (1-\bar v) p_{kR} ],
\end{equation}
where $\bar v = e\Vb/\go - \lfloor e\Vb/\go \rfloor $
is the fractional part of $e\Vb/\go$.
For $\bar v \ne 0,1$ there are 2 types of bidirectional processes
(labelled by L and R) with different number of attempts and
different probabilities of electron-hole pair creations.

We study bidirectional processes first in the regime $V_0=\Vb$ in
which the ac amplitude and the dc offset increase simultaneously.
The excess noise $S_I - S_{\rm dc}$ for different time-dependent
voltages is shown in Fig.~\ref{fig:ExcessNoiseVbEqV0}(a) for
comparison.
The probabilities $p_{kL(R)}$ as a function of voltage are shown
in Fig.~\ref{fig:ExcessNoiseVbEqV0}(b) for the cosine voltage
drive. The decomposition of the excess noise into contributions of
elementary processes is shown
in Fig.~\ref{fig:ExcessNoiseVbEqV0}(c). For the cosine voltage
drive there are 2 bidirectional processes (one of L-type and
another of R-type) which are excited per period. The L-type
processes transform continuously into the R-type ones at integer
values of $e\Vb/\go$, while the R-type processes disappear. The
step-like evolution of the probabilities of R-type processes as a
function of voltage does not introduce discontinuities in the
current noise because the corresponding number of attempts
vanishes. Instead, the interplay between L- and R-processes
results in kinks and the local minima at integer values of
$e\Vb/\go$, as shown in Fig.~\ref{fig:ExcessNoiseVbEqV0}(a). The
L-type (R-type) processes give the dominant contributions as
$e\Vb/\go$ approaches the integer values from the left (right)
because of the number of attempts which is proportional to $\bar v$
($1-\bar v$).

The Lorentzian voltage drive is special because it provides
the simplest one-particle charge transfer statistics. This is
achieved for impulses carrying an integer number of charge quanta
$eV_0/\go=N$ at offset voltages $\Vb \ge V_0$. In this case the
Lorentzian pulses do not create electron-hole pairs and only
single-particle processes given by Eq.~\eqref{eq:S2} remain. The
charge transfer statistics in each transport channel
is {\it exactly binomial}.%
\cite{art:LeeLevitovCONDMAT95,art:IvanovLeeLevitovPRB56-97} The
noise is reduced to the minimal noise level $S_{\rm dc}$ of the
effective dc bias, as shown in Fig.~\ref{fig:ExcessNoiseVbEqV0}(a).
A formal reason for this is the vanishing of the coefficients
$\fti_n=0$ for $n<-N$ [cf. Eqs.~\eqref{eq:SIT0-fn} and
\eqref{eq:fti-Res}]. A many-body state which is created by optimal
Lorentzian pulses has been obtained by Keeling {\it et
al.}\cite{art:KeelingKlichLevitovPRL97-06} recently.

\begin{figure}[t]
\includegraphics[scale=1.385]{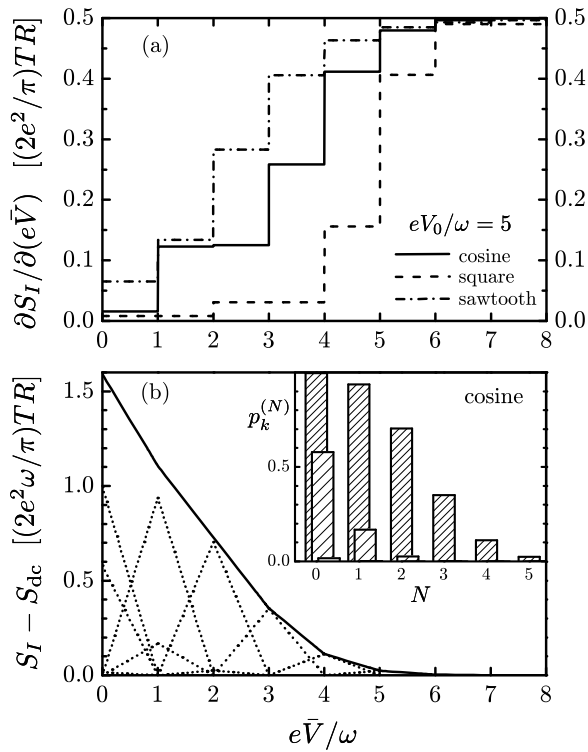}
\caption{\label{fig:LesovikLevitovPlots}
    The differential noise $\partial S_I/\partial(e\Vb)$
    as a function of dc offset $\Vb$ is shown in panel (a)
    for different time-dependent bias voltages: cosine (solid),
    square (dashed), and sawtooth (dash-dotted line).
    The ac amplitude of the drive is kept constant, $eV_0/\go = 5$.
    The excess noise $S_I-S_{\rm dc}$ as a function of $\Vb$
    (solid line) and the decomposition into contributions of
    elementary processes (dotted lines) are shown for the cosine
    drive in panel (b) [cf. Eq.~\eqref{eq:SI-Sdc}].
    The probabilities of elementary processes are shown in the
    inset.
}
\end{figure}

Now we focus on regime in which the ac component $\gD V(t)$ of the
drive is fixed and the dc offset $\Vb$ changes.
At low temperatures, the shot noise $S_I$ is a piecewise linear
function of the dc voltage offset $\Vb$ with kinks corresponding
to integer multiples of the driving frequency $e\Vb/\go = N$ and
slopes which depend on the shape and the amplitude of the ac
voltage component.%
\cite{art:LesovikLevitovPRL72-94,art:PedersenButtikerPRB58-98}
The differential noise $\partial S_I/\partial(e\Vb)$ consists of a
series of steps, as shown in
Fig.~\ref{fig:LesovikLevitovPlots}(a).
The piecewise linear dependence of noise can be understood in
terms of probabilities of elementary
processes.\cite{art:IvanovLevitovJETPLett58-93} For the ac
component fixed, the probabilities $p_{kR}$ and $p_{kL}$ of
electron-hole pair creations are {\it piecewise constant} as
a function of $\Vb$, and can be re-labelled by
$p_{kR}(\Vb) \equiv p_k^{(N)}$ and $p_{kL}(\Vb) \equiv p_k^{(N+1)}$
where $N=\lfloor e\Vb/\go \rfloor$.
This results in the piecewise linear dependence of the
excess noise given by Eq.~\eqref{eq:SI-Sdc} as a function of $\Vb$.
For the dc offset in the interval $m-1\le e\Vb/\go < m$ with
$m$ integer, the excess noise is a linear combination of
processes $p_k^{(m-1)}$ and $p_k^{(m)}$ ($k=1,2,\ldots$)
with the number of attempts proportional to $1-\bar v$ and $\bar
v$, respectively. For $m\le e\Vb/\go < m+1$ a different set of
processes $p_k^{(m)}$ and $p_k^{(m+1)}$ contributes and the excess
noise changes slope. The excess noise for harmonic drive is shown
in Fig.~\ref{fig:LesovikLevitovPlots}(b) (solid line) with
decomposition into contributions of elementary processes (dotted
lines). The probabilities of elementary processes are shown in the
inset.

\subsection{Beam splitter geometry}
\label{sec:BeamSplitter}

\begin{figure}[t]
\includegraphics[scale=0.65]{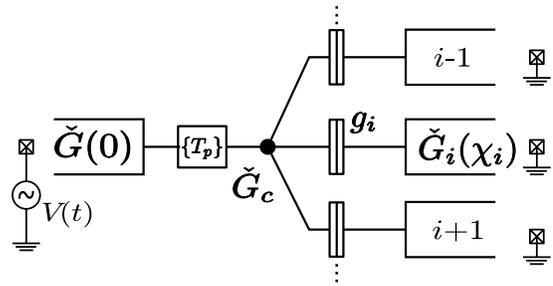}
\caption{\label{fig:MultitermCircuit1}
    Schematic representation of a multiterminal beam splitter.
    The source terminal is biased with a time-dependent voltage
    $V(t)$ with respect to the outgoing terminals. The total
    conductance $g_\gS=\sum_i g_i$ of the outgoing leads is
    assumed to be much larger than the conductance
    $g=(e^2/\pi)\sum_p T_p$ of the source contact.
}
\end{figure}

Here we study a multiterminal beam splitter geometry depicted in
Fig.~\ref{fig:MultitermCircuit1}.
The source terminal is biased with a time-dependent periodic voltage
$V(t)$ and through a mesoscopic conductor attached to several
outgoing terminals. The conductor is characterized by a set of
transmission eigenvalues $\{T_p\}$ and the outgoing leads
by conductances $g_i$. We are interested in the limit
in which the outgoing leads play a role of a detector and only
weakly perturb the charge transfer across the conductor. This is
achieved for the conductance $g_\gS\equiv \sum_i g_i$ to the
outgoing leads much larger than the conductance
$g\equiv (e^2/\pi)\sum_p T_p$ of a conductor. In this case the
particles which traverse the
conductor enter the outgoing terminals with negligible
backreflection into the source terminal. A similar setup with
spin-selective outgoing contacts has been used in
Ref.~\onlinecite{art:diLorenzoNazarovPRL94-05} to reveal singlet
electron states.

The cumulant generating function is calculated using the circuit
theory similarly as in Sec.~\ref{sec:CGF} for the 2-terminal case.
In contrast to Sec.~\ref{sec:CGF}, here we assign the counting
fields $\gx_i$ to the outgoing terminals. The Green's function
$\vG(0)$ of the source terminal and the Green's functions
$\vG_i(\gx_i)$ of the outgoing ones are given by
\begin{subequations}\label{eq:vG0vGi}
\begin{equation}
\vG(0) =    \begin{pmatrix}
            1 & 2\hti \\
            0 & -1
            \end{pmatrix}
\end{equation}
and
\begin{equation}
\vG_i(\gx_i) = e^{-i\gx_i \vgt_1/2}
            \begin{pmatrix}
            1 & 2h \\
            0 & -1
            \end{pmatrix} e^{i\gx_i \vgt_1/2}.
\end{equation}
\end{subequations}
Here $\hti$ and $h$ are the matrices in energy
indices defined in Sec.~\ref{sec:ElemProcTeq0-decomp}.
The Green's function
$\vG_c$ of the internal node is given by matrix current
conservation and normalization condition $\vG_c^2=\check 1$.
In the limit $g_\gS \gg g$, the central node is strongly
coupled to the outgoing terminals and the $\vG_c$ can be obtained
in the lowest order with the terminal $\vG$ unattached.
For simplicity we assume tunnel couplings to the outgoing
terminals. In this case matrix current conservation reduces
to\cite{art:NazarovSUPERLATTMICROSTRUCT25-99,art:BelzigCONDMAT02}
\begin{equation}\label{eq:vGcCurrconserv}
\left[\sum_i g_i \vG_i(\gx_i),\vG_c\right]=0.
\end{equation}

In the following we work in the low-temperature limit in which
$h^2=\hti^2=1$. We seek for the solution in the
form $\vG_c = \BF p_c\cdot \check{\pmb\tau}$ where
$\check{\pmb\tau}=(\vgt_1,\vgt_2,\vgt_3)$ is the vector of Pauli
matrices in Keldysh space.
From Eqs. \eqref{eq:vG0vGi} and \eqref{eq:vGcCurrconserv} we obtain
$\BF p_c = (h, ihc-s, ihs+c)$
where $c=\sum_i \tilde g_i \cos(\gx_i)$,
$s=\sum_i \tilde g_i \sin(\gx_i)$, and $\tilde g_i = g_i/g_\gS$.
The cumulant generating function of the charge transferred is
given by
\begin{equation}\label{eq:CGF-multitermBS}
\cS(\{\gx\}) = \sum_p \Tr\ln
\left[
    \check 1 + \frac{T_p}{2}
    \left(\frac{\aCom{\vG(0)}{\vG_c(\{\gx\})}}{2} - \check 1 \right)
\right],
\end{equation}
where summation over the internal Keldysh and energy indices is
assumed. The diagonalization of
$\aCom{\vG}{\vG_c}_{\cE'\cE''}$ is performed along the lines of
Ref.~\onlinecite{art:VanevicNazarovBelzigPRL99-07}.

For the cumulant generating function we obtain
$\cS(\{\gx\}) = \cS_{1L} + \cS_{1R}+\cS_2$
where $\cS_{1L,R}$ are the contributions of bidirectional processes
and $\cS_2$ is the contribution of the unidirectional ones.
The unidirectional processes are described by
\begin{equation}\label{eq:CGFS2-multitermBS}
\cS_2(\{\gx\}) = \frac{t_0e\Vb}{\pi} \sum_n
\ln \left( 1 + T_n \sum_i \tilde g_i(e^{i\gx_i}-1) \right),
\end{equation}
where $t_0$ is the measurement time and $\Vb$ is the dc voltage
offset (here we assume $e\Vb>0$).
The bidirectional processes are described by
\begin{multline}\label{eq:CGFS1LR-multitermBS}
\cS_{1\ga}(\{\gx\}) = M_\ga
\\
\times
\sum_n \sum_k \ln
\left[
1 + p_{k\ga} T_n R_n
\left(\sum_i \tilde g_i (e^{i\gx_i}+e^{-i\gx_i}-2) \right)\right.
\\
\left.
+
p_{k\ga} T_n^2 \left(\sum_{i<j} \tilde g_i\tilde g_j
(e^{i\gx_i}e^{-i\gx_j} + e^{-i\gx_i}e^{i\gx_j} - 2) \right)
\right].
\end{multline}
Here $k$ labels the bidirectional processes, $n$ labels transport
channels, $i$ and $j$ label the outgoing terminals, and $\ga=L,R$.
The number of attempts $M_\ga$ and the probabilities
$p_{k\ga}$ are the same as in Sec.~\ref{sec:ElemProcTeq0-decomp}.

Equations \eqref{eq:CGFS2-multitermBS} and
\eqref{eq:CGFS1LR-multitermBS} give the statistics of the charge
transfer in a multiterminal beam splitter at low temperatures
in the presence of a periodic time-dependent drive at the source
terminal. The $\cS_1(\{\gx\})$ and $\cS_2(\{\gx\})$ have a
direct physical interpretation.
The unidirectional processes, which are described by $\cS_2$, are
the single-electron transfers across the structure due to the dc
offset voltage $\Vb$ applied to the source terminal. The term
$T_n\tilde g_i e^{i\gx_i}$ in Eq.~\eqref{eq:CGFS2-multitermBS}
represents the process in which an electron in the $n$th
transport channel traverses the conductor with probability
$T_n$ and enters the outgoing terminal $i$ with probability
$\tilde g_i=g_i/g_\gS$.

On the other hand, the bidirectional processes represent the
electron-hole {\it pairs} created in the source terminal
and injected towards the conductor. The probabilities of such
excitations are given by $p_{k\ga}$. The interpretation of the
cumulant generating function given by
Eq.~\eqref{eq:CGFS1LR-multitermBS} can be obtained from a simple
counting argument. The term
$p_k T_n R_n \tilde g_i e^{i\gx_i}$ in
Eq.~\eqref{eq:CGFS1LR-multitermBS} represents the process in which
electron-hole excitation is created, hole is reflected,
and electron is transmitted into the outgoing terminal $i$.
Similarly, the term $p_k T_n R_n \tilde g_i e^{-i\gx_i}$
represents the process in which the electron is reflected
and the hole is transmitted. Finally, the term
$p_k T_n^2 \tilde g_i\tilde g_j e^{i\gx_i}e^{-i\gx_j}$
($i\ne j$) represents the process in which both particles are
transmitted with electron entering terminal $i$ and hole
entering terminal $j$.

Charge transfer statistics and current correlations
can be obtained using the cumulant generating function
$\cS(\{\gx\})$. For example, the current cross correlation between
{\it different} terminals $i$ and $j$ is given by
$S_{ij} = (e^2/t_0)\partial^2_{i\gx_i,i\gx_j} \cS|_{\gx=0}$:
\begin{equation}\label{eq:Sij-multiterminal}
S_{ij}=-\frac{2e^2\go}{\pi}\left(\sum_n T_n^2\right)
\tilde g_i \tilde g_j
\sum_k
[ \bar v  p_{kL} + (1-\bar v) p_{kR} ].
\end{equation}
We find that only bidirectional processes in which
both particles are transferred (one into the terminal $i$
and another into the terminal $j$) give the contributions to the
cross correlation $S_{ij}$. This has been obtained
previously by Rychkov {\it et al.}
\cite{art:RychkovPolianskiButtikerPRB72-05} in the special case
of harmonic drive in the limit of small driving amplitudes
($\bar V=0$ and $eV_0/\go \ll 1$).

The cross correlation $S_{ij}$ depends on bidirectional
processes and is proportional to the excess noise $S_I - S_{\rm dc}$
of a 2-terminal junction [Eq.~\eqref{eq:SI-Sdc}].
In a 2-termianal setup, the excess noise is just a small
correction to $S_{\rm dc}$ generated by unidirectional processes
for a bias voltage with $\Vb \simeq V_0$
(Fig.~\ref{fig:ExcessNoiseVbEqV0}).
The contribution of $S_{\rm dc}$ component can be reduced by
measuring current cross correlations between different outgoing
terminals in the beam splitter geometry with negligible
backscattering.

\section{Cumulants at $\Te \neq 0$}
\label{sec:CumulantsTneq0}

The full counting statistics and the corresponding elementary
transport processes obtained in Sec.~\ref{sec:ElemProcTeq0}
are valid description in the low temperature limit only, in which
electron-hole pairs are created by the applied voltage and no
thermally excited pairs exist. Formally, the diagonalization of
the operator in Eq.~\eqref{eq:Schi-h} in energy
indices, which is needed to deduce the elementary processes, is
based on the involution property of $h$-operators: $h^2=\hti^2=1$.
This property no longer holds at finite temperatures which are
comparable to the applied voltage. Nevertheless, the method we use
enables the efficient and systematic analytic
calculation of the higher-order cumulants at finite temperatures.

The cumulants can be obtained directly from Eq.~\eqref{eq:Schi-h} by
expansion in the counting field to the certain order
before taking the trace. The trace of a {\it finite number of
terms} can be taken in the original basis in which $h$ and
$\hti$ are defined. In the following we illustrate the approach
by calculation of the average current, the current noise power,
and the third cumulant at finite temperatures.
From Eq.~\eqref{eq:Schi-h} we obtain
\begin{subequations}\label{eq:d123Sdx123}
\begin{align}
\partial_{i\gx}\cS|_{\gx=0}
=&
\sum_p T_p \Tr_\cE(\hti-h),
\\
\partial^2_{i\gx}\cS|_{\gx=0}
=&
\sum_p \{
T_p \Tr_\cE (1-h\hti) - (T_p^2/2) \Tr_\cE [(h-\hti)^2] \},
\\
\partial^3_{i\gx}\cS|_{\gx=0}
=&
\sum_p
\{
(T_p^3/2) \Tr_\cE(\hti^3-h^3)
\notag
\\
&
+(3/2)T_p^2(1-T_p)\Tr_\cE [h\hti(\hti-h)]
\notag
\\
&
+T_p[1-(3T_p/2)] \Tr_\cE(\hti-h)\}.
\end{align}
\end{subequations}
In energy representation, the operators $h$ and $\hti$ are given by
\begin{subequations}\label{eq:hhtildeE1E2}
\begin{align}
h(\cE',\cE'')
=&
h(\cE')\; 2\pi\gd(\cE'-\cE'') \notag
\\
=&
\sum_{k,m} \fti_k\fti^*_{k+m} h(\cE')\;2\pi\gd(\cE''-\cE'-m\go),
\\
\hti(\cE',\cE'')
=&
\sum_{k,m} \fti_k \fti^*_{k+m} h(\cE'-k\go-e\bar V)
\notag
\\
&\times
\;2\pi\gd(\cE''-\cE'-m\go).
\end{align}
\end{subequations}
Here $h(\cE)=\tanh(\cE/2T_e)$ and we used the properties of
$\{\fti_n\}$ given by Eq.~\eqref{eq:fti-prop}.
After integration over energy in Eqs.~\eqref{eq:d123Sdx123}
we obtain the average current
$I=(e^2/\pi)\big(\sum_p T_p\big)\Vb$.
The current noise power
and the third cumulant are given by
\begin{multline}\label{eq:SIfiniteTemp}
S_I = \frac{e^2}{\pi}
\left[
2T_e \sum_p T_p^2   +  \sum_p T_p(1-T_p)
\right.
\\
\times
\left.
\sum_{n=-\infty}^\infty |\fti_n|^2
(e\Vb + n\go) \coth\left( \frac{e\Vb + n\go}{2T_e} \right)
\right]
\end{multline}
and
\begin{multline}
C_I = \frac{e^3}{\pi}
\bigg\{ e\Vb
\sum_p T_p(1-T_p^2)
+
3\sum_p T_p^2(1-T_p)
\\
\times
\sum_{n=-\infty}^\infty |\fti_n|^2
\left[
2T_e \coth\left( \frac{e\Vb + n\go}{2T_e} \right)
\right.
\\
\left.
-(e\Vb+n\go)\coth^2\left( \frac{e\Vb + n\go}{2T_e} \right)
\right]\bigg\},
\end{multline}
respectively. For a dc voltage bias $\fti_n = \gd_{n,0}$
and the noise and the third cumulant reduce to
$S_I = eFI$ and $C_I = e^2 F_3 I$ at zero temperature.
Here $F=[\sum_p T_p(1-T_p)]/(\sum_p T_p)$ is the Fano factor
and $F_3=[\sum_p T_p(1-T_p)(1-2T_p)]/(\sum_p T_p)$.

The average current is linear in dc voltage offset, which is
consistent with the initial assumption of energy-independent
transmission eigenvalues and instant scattering at the contact.
The result for the current noise power, Eq.~\eqref{eq:SIfiniteTemp},
describes the photon-assisted noise for arbitrary periodic voltage
drive. The coefficients $\fti_n$ for harmonic drive $V(t)=\Vb +
V_0 \cos(\go t)$ are given by the Bessel functions,
$\fti_n=J_n(eV_0/\go)$, and Eq.~\eqref{eq:SIfiniteTemp} reduces
to the previous results obtained by
Lesovik and Levitov\cite{art:LesovikLevitovPRL72-94} and
Pedersen and B{\"u}ttiker\cite{art:PedersenButtikerPRB58-98}
(see also Ref.~\onlinecite{art:BlanterButtikerPHYSREP336-00}).
The accurate noise measurements at
finite temperatures in the presence of the harmonic driving
are performed in
Ref.~\onlinecite{art:ReydelletRocheGlattliEtienneJinPRL90-03}.
The results are in agreement with Eq.~\eqref{eq:SIfiniteTemp}.

In the following we discuss the low- and high-temperature limits of
$S_I$ and $C_I$. At high temperatures
$T_e \gg |e\Vb|, n_0\go$, with $n_0\go$ being
the characteristic energy scale on which $|\fti_{n_0}|$ vanish,
the current noise power reduces to the thermal equilibrium
value $S_I = 2T_e G$, which is just a manifestation of the
fluctuation-dissipation theorem. The third cumulant is in this
regime proportional to the average current,
$C_I = e^2 F I$.
At high temperatures $S_I$ and $C_I$ carry no information on the
details of the time-dependent voltage drive.

At low temperatures $T_e \ll |e\Vb|, n_0\go$, the current noise
power reduces to
\begin{equation}\label{eq:SIT0-fn}
S_I = \frac{e^2}{\pi} \left( \sum_p T_p R_p \right)
\sum_{n=-\infty}^\infty |e\Vb + n\go|\; |\fti_n|^2.
\end{equation}
The differential noise $\partial S_I/\partial \Vb$ is a piecewise
constant function of $\Vb$ with steps
given by\cite{art:LesovikLevitovPRL72-94}
\begin{align}\label{eq:dSIdV-SlopeChange}
\left( \gD \frac{\partial S_I}{\partial\Vb}\right)_{e\Vb/\go=N}
&=
\frac{\partial S_I}{\partial\Vb}\bigg\vert_{N+0}
- \frac{\partial S_I}{\partial\Vb}\bigg\vert_{N-0} \notag
\\
&= \frac{2e^3}{\pi}\left( \sum_p T_pR_p\right) |\fti_{-N}|^2.
\end{align}
All steps add up to $(2e^3/\pi)\sum_p T_pR_p$ because of
$\sum_n |\fti_n|^2=1$, see Fig.~\ref{fig:LesovikLevitovPlots}(a).%
\cite{art:CommentPositive_eVbar}
The steps in $\partial S_I/\partial \Vb$
have been measured for harmonic drive in
normal\cite{art:SchoelkopfKozhevnikovProberRooksPRL80-98}
and normal-superconductor\cite{art:KozhevnikovProberPRL84-99}
junctions. In the superconducting state they appear at
integer values of $2e\Vb/\go$, which can be
interpreted as a signature of the elementary charge transport
processes in units of $e^*=2e$. The effective charge
is doubled in the superconducting state due to the Andreev
processes. We point out that for a general voltage drive, certain
steps at integer values of $e^*\Vb/\go=N$ may vanish if the
corresponding coefficient $\fti_{-N}=0$. For example, for a
square-shaped drive with integer amplitude $e^*V_0/\go=m$,
the steps at $e^*\Vb/\go = m+2k$ ($k\ne 0$) vanish
[cf. Eq.~\eqref{eq:fti-square} and
Fig.~\ref{fig:LesovikLevitovPlots}(a)].

At low temperatures, the third cumulant reduces to
$C_I = e^2 F_3 I$.
Unlike the current noise power, the third cumulant at low
temperatures does not depend on the ac component of the voltage
drive. This is because the bidirectional processes, created by the
ac voltage component, do not contribute to odd-order cumulants at
low temperatures [recall Eq.~\eqref{eq:S1LR}].

We conclude this section by comparison of two formulas for the
current noise power at zero temperature. For simplicity we take
$\Vb=0$. Equation~\eqref{eq:SIT0-fn} for the current noise power
reduces to
\begin{equation}\label{eq:SIT0Vb0-fn}
S_I = \frac{2e^2\go}{\pi} \left( \sum_p T_pR_p \right)
\left( \sum_{n=1}^\infty n |\fti_n|^2 \right).
\end{equation}
Here we used Eq.~\eqref{eq:fti-prop} to restrict the
summation to the positive $n$ only. On the other hand, the current
noise power is also given by Eq.~\eqref{eq:SI-ElemEvents}.
Regardless the similar form, the physical content of these two
equations is very different. Both equations give the same result
for $S_I$ as a consequence of the invariance of trace.
However, Eq.~\eqref{eq:SI-ElemEvents} has been obtained by taking
the trace of the operator in Eq.~\eqref{eq:Schi-h} in the basis
in which it is diagonal. 
The cumulant generating function given by Eq.~\eqref{eq:S1LR}
is decomposed into contributions of
elementary and statistically independent processes. The terms
proportional to $p_k$ which appear in Eq.~\eqref{eq:SI-ElemEvents}
are the contributions of these processes to the noise.
Equation~\eqref{eq:SIT0Vb0-fn} has been
obtained by taking the trace of the operator in Eq.~\eqref{eq:Schi-h}
in a basis in which it is not diagonal. Although the end result for
$S_I$ is the same, the individual terms proportional to
$n|\fti_n|^2$ which appear in Eq.~\eqref{eq:SIT0Vb0-fn} have no
direct physical interpretation.

In the limit of small-amplitude voltage drive, $eV_0/\go \ll 1$,
only one electron-hole pair is excited per period with probability
$p_1$. Comparing Eqs.~\eqref{eq:SI-ElemEvents} and
\eqref{eq:SIT0Vb0-fn} we obtain that in this case
$p_1 = \sum_{n=1}^\infty n |\fti_n|^2$.
In fact, as shown in Fig.~\ref{fig:ElemProcCos} for harmonic drive,
the assumption of small amplitudes can be relaxed to the
amplitudes comparable or even larger than the drive frequency.
The accuracy of this approximation depends on the actual
voltage drive (cf. Figs.~\ref{fig:ElemProcCos}
and \ref{fig:NoiseDerivVbEq0All}).

\section{Charge transfer processes at $\Te \neq 0$}
\label{sec:ElemProcTne0}

Here we study the effect of finite temperature on elementary
charge transfer processes.
For a periodic voltage drive, the
cumulant generating function $\cS(\gx)$ in
Eq.~\eqref{eq:Schi-h} can be recast in the following form:
\begin{equation}\label{eq:CGFfiniteTemp}
\cS(\gx) = \frac{t_0}{\pi}
\sum_p \int_0^\go d\cE\;
\ln \left(
    1+\sum_{n=-\infty}^\infty q_n (e^{in\gx}-1)
    \right).
\end{equation}
Here $q_n e^{in\gx}$ describes the process in which $n$ charges
are transmitted through the scatterer with probability $q_n$. The
charge transfer can occur in either direction depending on the
sign of $n$. To obtain the probabilities $q_n$ we diagonalize the
operator under the logarithm in Eq.~\eqref{eq:Schi-h} numerically,
multiply all the eigenvalues, and take the inverse Fourier
transformation with respect to $\gx$. For simplicity we focus on
ac voltage drive with no dc offset. For a given ac drive, the
probabilities $q_n$ depend on energy, temperature, and transmission,
$q_n\equiv q_n(\cE/\go, \Te/\go; T)$. Transport properties of
the junction enter through summation over the transmission
eigenchannels $\{T_p\}$.

Equation \eqref{eq:CGFfiniteTemp} allows us to study the crossover
from ac-driven fluctuations at zero temperature to thermal
fluctuations at temperatures much larger than the voltage drive.
The important difference between the two regimes is that ac drive
mixes electron states of different energies in contrast to
thermal fluctuations which are diagonal in energy. As
we have seen, this is also reflected in the charge transfer
statistics. The statistics for an ac drive (at zero temperature)
can be interpreted in terms of electron-hole pairs after mapping
the problem into energy interval set by the driving frequency
$\go$. The mixing of energy states is taken into account by
diagonalization of the remaining matrix structure in
energy. On the other hand, the statistics in the thermal limit is
composed of electron transfers (in either direction) which are
independent at different energies. In the crossover regime there
is still some mixing of different energy states present and we use
the probabilities $q_n$ in Eq.~\eqref{eq:CGFfiniteTemp} with energy
mapped in the $\go$-interval. These different physical situations
are depicted schematically in Fig.~\ref{fig:ElemProcDiagram}.

\begin{figure}[t]
\includegraphics[scale=1.2]{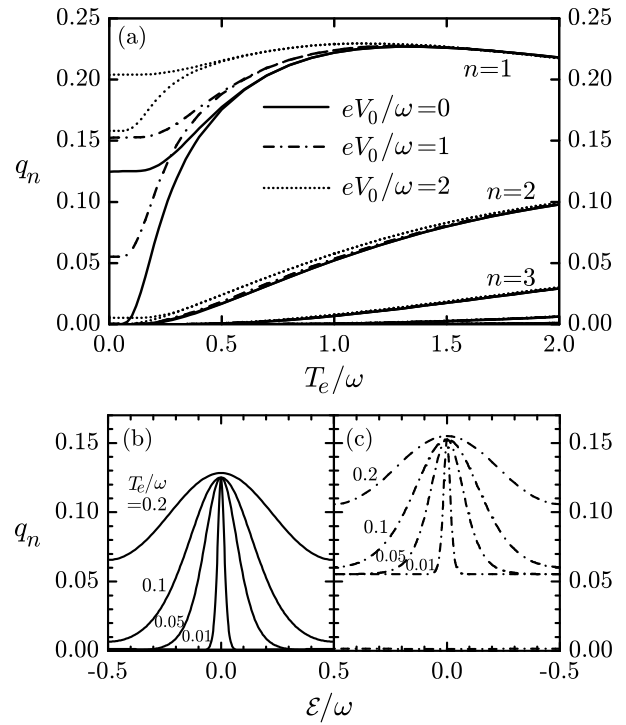}
\caption{\label{fig:probVSTempEt05}
    The probabilities $q_{\pm n}$ of multiple charge transfers for
    harmonic voltage drive $V(t)=V_0\cos(\go t)$ and transmission
    $T_p=0.5$. The temperature dependence of $q_n$ is shown in
    panel (a) for different driving amplitudes. The energy
    dependence of $q_n$ at low temperatures is indicated by plots
    for $\cE/\go=0$ (top line) and $\cE/\go=0.5$ (bottom line).
    The energy dependence of $q_n$ is shown in the thermal
    limit $eV_0/\go = 0$ (b) and for $eV_0/\go = 1$ (c). At zero
    temperature $q_{\pm 1}$ do not depend on energy and are given
    by $q_{\pm 1}=T(1-T)p_1(eV_0/\go)$ (cf.
    Fig.~\ref{fig:ElemProcCos}). As the temperature increases,
    multiple charge transfers start to enter the transport.
    The limit of thermal fluctuations is reached at temperatures
    larger than the driving amplitude. It can be recast
    again into single-charge transfers independent in the full
    energy range.
}
\end{figure}

The probabilities $q_n$ are shown in Fig.~\ref{fig:probVSTempEt05}
for cosine voltage drive $V(t)=V_0\cos(\go t)$ and a transport
channel of transmission $T_p=0.5$.
We first discuss thermal limit of probabilities $q_n$
shown by solid lines in Figs.~\ref{fig:probVSTempEt05} (a), (b).
For $\go \gg \Te$, the mapping into interval $\go$ is irrelevant
since it is larger than the energy scale of thermal fluctuations.
In this case the probabilities reduce to $q_{\pm 1}=f(1-f)T_p$ and
$q_n=0$ otherwise, in accordance with Eq.~\eqref{eq:CGFdc}
[cf. Fig.~\ref{fig:probVSTempEt05} (b)]. As the mapping interval
becomes comparable or smaller than the temperature, the
higher-order ``bands'' $q_n$ ($n=\pm 1,\pm 2,\ldots$) start to
appear. For $\go \lesssim 2 \Te$ the probabilities $q_n$ are
independent of energy
[Fig.~\ref{fig:probVSTempEt05} (a)] and are given by
\begin{equation}\label{eq:qn}
q_n = \int_{-\pi}^\pi \frac{d\gx}{2\pi}\;
e^{\tilde\cS(\gx)-i n \gx}
\end{equation}
with 
$\tilde \cS(\gx) = - 4(\Te/\go) \arcsin^2[\sqrt{T_p} \sin(\gx/2)]$.
Here $\tilde \cS(\gx)$ is related to the cumulant generating
function $\cS(\gx)$ in thermal equilibrium by
$\cS(\gx)=(2t_0/\gt)\sum_p \tilde \cS(\gx)$. The latter is
obtained from Eq.~\eqref{eq:CGFdc} after energy
integration.\cite{art:LevitovLeeLesovikJMathPhys37-96}

We emphasize again that the picture of multiple charge transfers
{\it in thermal equilibrium limit} is solely due to mapping into
the energy interval $\go$ in Eq.~\eqref{eq:CGFfiniteTemp}.
Although such mapping is not needed because thermal fluctuations
are diagonal in energy, we nevertheless perform it here to
provide a limit for ac-driven processes as the temperature
increases. At temperatures much larger than ac drive there is no
mixing of different energy states and the statistics in the full
energy range reduces to single electron transfers which are
independent at different energies [Eq.~\eqref{eq:CGFdc}].

Now we focus on the ac-driven processes and the crossover region
between ac and thermal fluctuations as the temperature increases.
For simplicity, we consider driving
amplitudes $eV_0/\go=1$ and $eV_0/\go=2$.
In this case $q_{\pm 1}(T_p)=T_p(1-T_p) p_1(eV_0/\go)$
at zero temperature and do not depend on energy.
These single-charge transfers originate from an
electron-hole pair which is created with probability $p_1$ per
voltage cycle (Fig.~\ref{fig:ElemProcCos}).
As the temperature increases, the interplay of thermal and ac
excitations introduces a nontrivial energy dependence of
probabilities $q_{\pm 1}$ shown in Fig.~\ref{fig:probVSTempEt05} (c).
It also modifies the dependence on transmission eigenvalues with
$q_{\pm 1}(T_p)$ no longer being proportional to $T_p(1-T_p)$
[cf. Eq.~\eqref{eq:qn}].
As the temperature increases further, the multiple-charge
transfers come into play as shown in
Fig.~\ref{fig:probVSTempEt05} (a).
At temperatures larger than the driving amplitude the limit of
thermal transport is reached with statistics which can be
recast again in the form of single-charge transfers, independent at
different energies.


\section{Conclusion}
\label{sec:Conclusion}

We have studied charge transfer statistics in a generic mesoscopic
contact driven by a time-dependent voltage. We have obtained the
analytic form of the cumulant generating function at zero
temperature and identified the elementary charge transfer
processes. The unidirectional processes represent electrons which
are injected from the source terminal due to excess dc bias
voltage. The bidirectional processes represent electron-hole pairs
which are created by the time-dependent voltage bias and injected
towards the contact. This interpretation is consistent with the
charge transfer statistics in a multiterminal beam splitter
geometry in which electrons and holes can be partitioned into
different outgoing terminals.


The elementary charge transfer processes can be probed by current
noise power and higher-order current correlators. The
bidirectional processes contribute to the noise and higher-order
even cumulants of the transferred charge at low temperature and
give no contributions to the average current and higher-order odd
cumulants. For an ac voltage with no dc offset, the noise is
entirely due to bidirectional processes. The individual processes
can be identified from the oscillations of the differential noise
$\partial S_I/\partial V_0$ as the amplitude $V_0$ of the drive
increases.


A time-dependent voltage drive with a nonzero dc offset generates
both unidirectional and bidirectional processes. The bidirectional
processes give rise to the excess noise with respect to dc noise
level which is set by the unidirectional processes. The excess noise
can be probed by measuring current cross correlations between
different outgoing terminals in the beam splitter geometry. The
cross correlations are only due to processes in which the incoming
electron-hole pair is split and the particles enter different
outgoing terminals.


The method we use enables the systematic calculation of the
higher-order current correlators at finite temperatures by
expansion of the cumulant generating function in the counting
field. We have obtained the current noise power and the third
cumulant for arbitrary periodic voltage applied.

We have also studied the effect of finite temperature on
elementary charge transfer processes. In the limits of low (high)
temperatures the single-charge transfers occur in either direction
due to ac excited (thermally excited) electron-hole pairs.
However, the nature of the two is very different: the ac drive
mixes the electron states of different energies while
thermal fluctuations are diagonal in energy. In the crossover
region there is still some mixing of different energy states
present. Such an interplay of thermal and ac excitations can be
interpreted in terms of multiple-charge transfers with
probabilities dependent on energy and temperature.

\section*{ACKNOWLEDGMENTS}

This work has been supported by the German Research Foundation
(DFG) through SFB 513 and SFB 767 and the Swiss National Science
Foundation (SNSF).

\section*{APPENDIX}

\subsection{Determinants of block matrices}

Let $\BF A$, $\BF B$, $\BF C$, and $\BF D$ be the quadratic
matrices of the same size. Then the following equalities hold:
\begin{equation}\label{eq:DetBlockMatr}
\det \begin{pmatrix}
     \BF A & \BF B \\
     \BF C & \BF D
     \end{pmatrix}
=
\begin{cases}
\det(\BF A\BF D - \BF B\BF C), & [\BF C, \BF D]=0, \\
\det(\BF D\BF A - \BF B\BF C), & [\BF B, \BF D]=0, \\
\det(\BF D\BF A - \BF C\BF B), & [\BF A, \BF B]=0, \\
\det(\BF A\BF D - \BF C\BF B), & [\BF A, \BF C]=0.
\end{cases}
\end{equation}
In the case in which more than two blocks commute with each other,
the corresponding determinants on the right hand side of
Eq.~\eqref{eq:DetBlockMatr} coincide.

\subsection{Coefficients $\fti_n$}
\label{sec:fti-calculation}

Here we calculate coefficients $\{\fti_n\}$ given by
Eq.~\eqref{eq:fti-def} for different time-dependent voltages.
For the cosine voltage drive $V(t)=\Vb + V_0\cos(\go t)$,
the coefficients $\fti_n$ can be calculated using Jacobi-Anger
expansion:\cite{art:AbramovicStegun}
\begin{equation}
e^{i z \sin(\theta)} = \sum_{n=-\infty}^\infty J_n(z)\;e^{in\theta},
\end{equation}
where $J_n$ are the Bessel functions of the first kind.
From Eq.~\eqref{eq:fti-def} we obtain
\begin{equation}
\fti_n = J_n(eV_0/\go).
\end{equation}

The square voltage drive is given by $\gD V(t)=V_0$ for $0<t<\gt/2$
and $\gD V(t)=-V_0$ for $\gt/2<t<\gt$. For noninteger values of
$eV_0/\go$, the coefficients $\fti_n$ are given by
\begin{equation}\label{eq:fti-square}
\fti_n = \frac{2}{\pi}\frac{eV_0}{\go}
\frac{\sin[(n-eV_0/\go)\pi/2]}
{n^2 - (eV_0/\go)^2}\;
e^{i (n - eV_0/\go) \pi/2}.
\end{equation}
For integer values of $eV_0/\go$, the coefficients $\fti_n$ are
obtained by taking the limit in the previous formula.

The sawtooth voltage drive is given by $\gD V(t)=2V_0t/\gt - V_0$
for $0<t<\gt$. In this case
\begin{multline}
\fti_n = \frac{1}{2\sqrt{2eV_0/\go}}
\exp
\left( i\frac{\pi}{2}\frac{(eV_0/\go+n)^2-eV_0/2\go}{eV_0/\go}
\right)
\\
\times
\left[
\erf
    \left( \frac{\sqrt{\pi}(eV_0/\go-n)}{\sqrt{2eV_0/\go}}e^{i\pi/4}
    \right) \right.
\\
+
\erf
\left.
    \left( \frac{\sqrt{\pi}(eV_0/\go+n)}{\sqrt{2eV_0/\go}}e^{i\pi/4}
    \right)
\right],
\end{multline}
where $\erf(z)=(2/\sqrt{\pi})\int_0^z dt\; \exp(-t^2)$ is the
error function.

The triangle voltage drive is characterized by
$\gD V(t)=4V_0t/\gt-V_0$ for $0<t<\gt/2$ and
$\gD V(t)=-4V_0t/\gt+3V_0$ for $\gt/2<t<\gt$. In this case
\begin{multline}
\fti_n = \frac{1}{2\sqrt{eV_0/\go}}
\Re
\left\{
\exp
    \left( i\frac{\pi}{4}\frac{(eV_0/\go+n)^2-eV_0/\go}{eV_0/\go}
    \right)
\right.
\\
\times
\left[
\erf\left( \frac{\sqrt{\pi}(eV_0/\go-n)}{2\sqrt{eV_0/\go}}e^{i\pi/4}
    \right)
\right.
\\
+
\left.\left.
\erf\left( \frac{\sqrt{\pi}(eV_0/\go+n)}{2\sqrt{eV_0/\go}}e^{i\pi/4}
    \right)
\right]\right\}.
\end{multline}

The voltage drive which consists of Lorentzian voltage pulses of
width $\gt_L$ is given by
\begin{align}
\gD V(t)
&=
-V_0 + \frac{V_0}{\pi} \sum_{k=-\infty}^\infty
\frac{\gt\gt_L}{(t-k\gt)^2+\gt_L^2} \notag
\\
&=
-V_0 +
\frac{V_0 \sinh(2\pi\gt_L/\gt)}
{\cosh(2\pi\gt_L/\gt)-\cos(2\pi t/\gt)}.
\end{align}
The total voltage $V(t)=\Vb+\gD V(t)$ varies between
$V_{\rm min}=\Vb+V_0[\tanh(\pi\gt_L/\gt)-1]$ and
$V_{\rm max}=\Vb + V_0[\cotanh(\pi\gt_L/\gt)-1]$. The coefficients
$\fti_n$ are given by
\begin{multline}
\fti_n = e^{-i\pi eV_0/\go}
\int_{-1/2}^{1/2}dx \; \frac{\{\sin[\pi(x+iy)]\}^{eV_0/\go}}
{\{\sin[\pi(x-iy)]\}^{eV_0/\go}}
\\
\times
e^{i2\pi(eV_0/\go+n)x},
\end{multline}
where $y=\gt_L/\gt$. For integer values $eV_0/\go=N>0$ we obtain
the simplified expressions:
\begin{equation}\label{eq:fti-Res}
\fti_n = (-1)^N 2\pi i
\Res_{x=iy}
\left(
\frac{\sin^N [\pi(x+iy)]}{\sin^N [\pi(x-iy)]}
\; e^{i2\pi(n+N)x}
\right)
\end{equation}
for $n>-N$, $\fti_n=0$ for $n<-N$, and
$\fti_{n=-N}=(-1)^N e^{-2\pi Ny}$. In particular, the coefficients
$\fti_n$ for $eV_0/\go=1$ are given by
$\fti_n = e^{-2\pi ny} - e^{-2\pi(n+2)y}$ for $n>-1$,
$\fti_{-1}=-e^{-2\pi y}$, and $\fti_n=0$ otherwise.
For $eV_0/\go=2$ we obtain
$\fti_n = e^{-2\pi ny}(1-e^{-4\pi y})[n+1-(n+3)e^{-4\pi y}]$
for $n>-2$, $\fti_{-2}=e^{-4\pi y}$, and $\fti_n=0$ otherwise.

\bibliography{bibliography}

\begin{thebibliography}{63}
\bibitem[{art({\natexlab{d}})}]{art:CommentCGFeq}
\bibinfo{note}{Equation~\eqref{eq:Schi-h} coincides with Eq.~(5) of
  Ref.~\onlinecite{art:VanevicNazarovBelzigPRL99-07} which has been obtained by
  a different method. The two terms in Eq.~(5) of
  Ref.~\onlinecite{art:VanevicNazarovBelzigPRL99-07} give the same
  contributions, in agreement with Eq.~\eqref{eq:LevitovCGF2terminal} of the
  present manuscript.}
\bibitem[{art({\natexlab{e}})}]{art:SIconvention}
\bibinfo{note}{We use the following convention for the current correlation
  function: $S_I(t'-t'')=(1/2) \langle \aCom{\gD \hat I(t')}{\gD \hat
  I(t'')}\rangle$, where $\gD \hat I(t) = \hat I(t) - \langle \hat I \rangle$.
  The current noise power $S_I \equiv S_I(\Omega = 0)$ is the zero-frequency
  component of the Fourier transform $S_I(\Omega)$.}
\bibitem[{art({\natexlab{f}})}]{art:CommentPositive_eVbar}
\bibinfo{note}{The differential noise is an odd function of $\Vb$ for the
  voltages considered in Fig.~\ref{fig:LesovikLevitovPlots} and is shown for
  $e\Vb>0$ only.}
\end{thebibliography}

\end{document}